\documentclass[final]{pasj00}

\SetRunningHead{Aoki et al.}{AKARI~J1757+5907}
\Received{2010/09/10}
\Accepted{2011/01/14}
\Published{vol. 63, Subaru special issue}

\begin{document}

\title{Outflow in Overlooked Luminous Quasar: 
Subaru Observations of AKARI~J1757+5907
\thanks{Based on data collected at Subaru Telescope, which is operated by the National Astronomical Observatory of Japan.}}

\author{Kentaro \textsc{Aoki},\altaffilmark{1}
Shinki \textsc{Oyabu},\altaffilmark{1,2} 
Jay~P. \textsc{Dunn},\altaffilmark{3} 
Nahum \textsc{Arav},\altaffilmark{3} 
Doug \textsc{Edmonds},\altaffilmark{3} 
Kirk~T. \textsc{Korista},\altaffilmark{4} 
Hideo \textsc{Matsuhara},\altaffilmark{5} 
\and 
Yoshiki \textsc{Toba}\altaffilmark{5} 
}
\altaffiltext{1}{Subaru Telescope, National Astronomical Observatory of Japan,
    650 North A'ohoku Place, Hilo, \\HI 96720, U.S.A.}
\altaffiltext{2}{Graduate School of Science, Nagoya University, Chikusa-ku, Nagoya 464-8602}
\altaffiltext{3}{Department of Physics, Virginia Tech, Blacksburg, VA 24061, U.S.A.}
\altaffiltext{4}{Department of Physics, Western Michigan University, Kalamazoo, MI 49008-5252, U.S.A.}
\altaffiltext{5}{Institute of Space and Astronautical Science, Japan Aerospace Exploration Agency, 3-1-1 Yoshinodai, Sagamihara, Kanagawa 229-8510 }

\maketitle

\begin{abstract}
We present Subaru observations of the newly discovered luminous quasar
AKARI~J1757+5907, which shows an absorption outflow in its spectrum.
The absorption consists of 9 distinct troughs, and our analysis
focuses on the troughs at $\sim -1000$ km s$^{-1}$ for which we can
measure accurate column densities of He~\emissiontype{I}*,
Fe~\emissiontype{II} and Mg~\emissiontype{II}.  We use photoionization
models to constrain the ionization parameter, total hydrogen column
density, and the number density of the outflowing gas.  These
constraints yield lower limits for the distance, mass flow rate and
kinetic luminosity for the outflow of 3.7 kpc, $70~M_{\odot}$
yr$^{-1}$, and $2.0 \times$10$^{43}$ ergs s$^{-1}$, respectively. Such
mass flow rate value can contribute significantly to the metal
enrichment of the intra-cluster medium.  We find that this moderate
velocity outflow is similar to those recently discovered in massive
post-starburst galaxies.  Finally, we describe the scientific potential
of future observations targeting this object.
\end{abstract}

\section{Introduction}
Roughly 20\% of all quasars exhibit Broad Absorption Lines (BAL) in
their spectrum \citep{2008MNRAS.386.1426K,2003AJ....125.1784H}, which
are indicative of outflows with velocities $\sim10^3-10^4$ km s$^{-1}$.
Kinetic energy and mass emanating from quasars have become 
key elements in theoretical modeling of the evolution of supermassive
black holes (SMBH) and their host galaxies (e.g.,
\cite{2005Natur.433..604D,2008ApJS..175..356H}), the suppression of
cooling flows in clusters (e.g.,
\cite{2010ApJ...717..708C,2009MNRAS.398..548B}), and enrichment of the
intra-cluster and inter-galactic media with metals (e.g.,
\cite{2007AA...463..513M}).  Collectively, harnessing quasars'
mechanical energy to help in driving the above processes is known as
``AGN feedback''. 
 \par To assess the contribution of BAL outflows to AGN feedback scenarios, it is (at least)
necessary to determine their  mass
flow rate ($\dot{M}_{out}$) and kinetic luminosity ($\dot{E}_k$). Early attempts to do so were done by
\citet{2001ApJ...548..609D}, \citet{2002ApJ...570..514D},
\citet{2001ApJ...550..142H}, and \citet{1995ApJ...443..586W}.
Recently, using improvements in analysis methods and in target
selection (see discussions in
\cite{2008ApJ...681..954A,2010ApJ...709..611D}), we published several
more accurate determinations of these quantities (Moe et al. 2009;
Dunn et al. 2010; Bautista et al 2010; Arav et al 2010).  Here we present a similar
analysis of an outflow in a luminous overlooked quasar.  
\par We
use the term ``BAL outflow'' to designate intrinsic absorption
detected in the spectrum of a quasar (i.e., originating from
outflowing material in the vicinity of the AGN, see
\cite{1997AJ....113..136B}). The original BAL definition
\citep{1991ApJ...373...23W} was created to differentiate, in
low-resolution spectra, AGN outflow absorption systems from
intervening absorbers that do not have a dynamical connection to the
AGN and is now physically obsolete. 
Significant number of narrower absorption lines show intrinsic natures, i.e.,  
time variability and partial coverage of  the background light source.
The frequency of intrinsic absorption lines are discussed and summarized in \citet{2008ApJ...672..102G}.
We therefore use ``BAL outflow''
to designate the physical nature, rather than the observational
definition, of the phenomenon.  
\par AKARI-IRC-V1~J1757000+590759
(hereafter AKARI~J1757+5907) was discovered during the follow-up
observations of AKARI mid-infrared (MIR) All-Sky Survey.  The infrared
satellite AKARI performed an all-sky survey at 9 and 18
$\mu\mathrm{m}$ as well as at four far-infrared bands
\citep{Murakami07,Ishihara2010}.  The initial identification of the
AKARI MIR All-Sky Survey sources involved association with the Two
Micron All Sky Survey (2MASS) catalog \citep{skrutskie06}.  This
search identified some AKARI MIR sources with $F(9\mu\mathrm{m})/F(Ks)
> 2$ in the high galactic latitude ($|b|>20^{\circ}$) after excluding the sample
in the Large and Small Magellanic Clouds.  AKARI~J1757+5907 has a
large ratio of mid-IR to near-IR flux density
($F(9\mu\mathrm{m})/F(Ks) =11.1$).  This mid-IR source is also
coincident with a bright near-UV and optical source.  We present
photometries of AKARI~J1757+5907 in table~\ref{tbl1}.  The
photographic magnitudes are from USNO-B1.0 catalog \citep{Monet03},
and they are converted to $g, r,$ and $i$ magnitude using equation 2
of \citet{Monet03}.
We found the radio source NVSS~J175659.82+590801.5 ($6.5 \pm 0.5$ mJy at 1.4 GHz) \citep{1998AJ....115.1693C}
is coincident with AKARI~J1757+5907.
The ratio of radio (5 GHz) to optical (4400 {\AA}) flux density is 1.4 assuming 
radio spectral index is -0.5.
The ratio indicates the quasar is radio quiet 
\citep{1989AJ.....98.1195K}.
\par
The follow-up spectroscopy using KPNO 2.1m
telescope revealed that AKARI~J1757+5907 is a $z=0.615$ quasar that
shows He~\emissiontype{I}* and Mg~\emissiontype{II} absorption lines
as well as H$\beta$ and strong Fe~\emissiontype{II} emission lines
(Toba~et~al, in preparation).  The spectrum resembles the one of QSO~2359-1241
\citep{Brotherton2001, Arav2001}.
Both quasars show rare He~\emissiontype{I}* absorption as well as
Mg~\emissiontype{II} absorption.
They also have redder continua and strong Fe~\emissiontype{II} emission lines.
The high resolution spectroscopy of QSO~2359-1241 by \citet{Arav2001}
revealed Fe~\emissiontype{II} absorption lines,
Thus Fe~\emissiontype{II} absorption lines are
expected to exist in AKARI~J1757+5907.
Following \citet{Korista2008}, by measuring the ionic column densities
($N_{ion}$) of Fe~\emissiontype{II} and He~\emissiontype{I}*, we can
constrain the ionization parameter ($U_{H}$) and hydrogen column
density ($N_{H}$).  The high brightness of this quasar permits us to
do high resolution spectroscopy of He~\emissiontype{I}* absorption
lines and search for Fe~\emissiontype{II} resonance and excited state
absorption lines.  The $N_{ion}$ ratio of Fe~\emissiontype{II}* to Fe~\emissiontype{II}(E=0) 
yields the hydrogen number density, which in turn yields the
 distance of outflowing gas from the central
source, $\dot{E}_k$ and $\dot{M}_{out}$ by using the number density,
$U_H$ and $N_H$.  
\par
The plan of this paper is as follows.  In \S~2 we describe the
observations and data reduction.  In \S~3 we determine the redshift of
the object. The outflow absorption troughs are discussed in \S~4 and
the spectral energy distribution in \S~5. In \S~6 we describe our
photoionization modeling, and in \S~7 the resultant determination of
mass flow rate and kinetic luminosity for the outflow. In \S~8 we
discuss our results and in \S~9 we describe the scientific potential
of additional Subaru/HDS and HST/COS observations targeting this
object.
Throughout this paper we assume $H_{\rm 0} = 70$ km s$^{-1}$ Mpc$^{-1}$,
$\Omega_{m} = 0.3$, and $\Omega_{\Lambda} = 0.7$.  Note that
wavelengths of any transition in this paper are ones in vacuum, and
the observed wavelength scale is converted to one in vacuum.
\section{Observations and Data Reduction} \label{Obs}
\subsection{high resolution spectroscopy}
The high resolution spectroscopy of AKARI~J1757+5907 
(R.A.=17:57:00.24, Decl.=+59:08:00.3 (J2000.0)) was done
with the High Dispersion Spectrograph
(HDS; \cite{Noguchi2002})
attached to the Subaru 8.2 m telescope \citep{Iye04} on 2010 June 17 (UT). 
The weather conditions were poor with thick cirrus and the seeing was unstable ($> $\timeform{1.0''}). 
The slit width was set to be \timeform{1.0''}. 
The HDS setting was Yd covering between 4054 and 6696 {\AA}. 
This results in a resolving power of $R\sim 36000$.
The binning was 2 (spatial direction) $\times$ 4 (dispersion direction).
We obtained eight exposures of 1800 s each, however,
one of them is unusable due to low signal-to-noise ratio.
\par
The data were reduced using IRAF\footnote{IRAF is distributed by the National 
Optical Astronomy Observatory, which is operated by the Association of 
Universities for Research in Astronomy, Inc., under cooperative agreement with 
the National Science Foundation.} 
for the standard procedures of overscan subtraction,
dark subtraction, cosmic ray removal and flat-fielding, where
wavelength calibration was performed using the Th-Ar lamp.
the rms wavelength calibration error is 0.011 - 0.013 {\AA}.
The one-dimensional spectra were extracted from each exposure.
Heliocentric correction was applied.
After that, all the spectral exposures  were combined, and all orders were connected
to one spectrum. 
We normalized the spectrum using spline fits.
Finally we converted to vacuum wavelength scale.

\subsection{Spectrophotometry}
The low resolution spectrophotometry of AKARI~J1757+5907 was done
on 2010 June 30 (UT) with
FOCAS \citep{Kashik02} attached to the Subaru telescope.
The new fully-depleted-type CCDs developed by NAOJ/ATC and fabricated by Hamamatsu Photonics K. K. were installed and commissioned at that time.
We obtained six spectra of 5 minutes integration under a clear condition and 
good seeing (\timeform{0.6''} - \timeform{0.9''} ).  
The \timeform{2.0''} width slit was used for spectrophotometry purpose. 
We used two configurations: the R300 grism with the O58 filter (`red') and the B300 grism without any order-cut filters (`blue').
The first three 300s spectra were the `red' which
covers between 5700 {\AA} and 10200 {\AA}.  
The last three 300s spectra were the `blue' which have an uncontaminated range
between 3500 {\AA} and 7000 {\AA}.   
The atmospheric dispersion corrector was used.
The slit position angle was \timeform{0D}, and the binning was 2 (spatial direction) $\times$ 1 (dispersion direction)  
The spectrophotometric
standard star BD$+28^{\circ}4211$ was observed for sensitivity calibration.
\par
The data were reduced using IRAF
for the standard procedures of bias subtraction, wavelength
calibration, and sky subtraction, except for flat-fielding.  
AKARI~J1757+5907 is so bright that its counts on the CCD are comparable to the flat frames, and are much higher at shorter wavelength ($< 4000$ {\AA}).
The flat-fielding procedure significantly reduced its signal-to-noise ratio.
We therefore skipped the flat-fielding procedure.
Wavelength calibration was
performed using OH night sky emission lines for the red spectra
and the Cu-Ar lamp for the blue ones.
The rms error of wavelength calibration is 0.2 {\AA}.  
The seeing was much smaller than the slit width, 
thus the resolution was determined by the seeing disk size.
The resulting He~\emissiontype{I}* 3889 absorption line is 13 {\AA} width at 6260 {\AA}.
This value corresponds to the resolving power of 480 ($\sim 620$ km s$^{-1}$),  
which is similar to the resolution obtained with the \timeform{0.8"} width slit,
and is consistent with the seeing size during our observations.
The sensitivity calibration was
performed as a function of wavelength.
The flux of the blue and red spectra at the same wavelength agree within 1.6\%.
The foreground Galactic
extinction of $E(B-V)=0.043$ mag \citep{SFD98} was corrected.
\par
\section{Results of spectrophotometry} \label{Results}
\begin{figure}
  \begin{center}
    \FigureFile(90mm,110mm){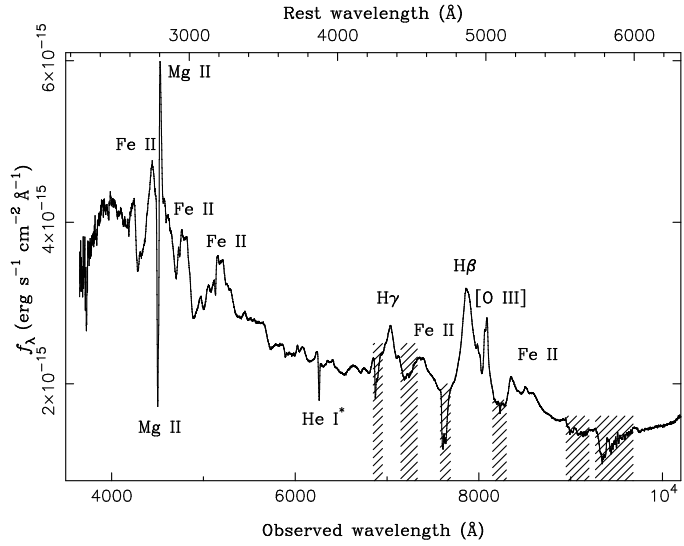}
  \end{center}
  \caption{Low resolution spectrum of AKARI~J1757+5907.
Ordinate is a flux density corrected for the Milky Way extinction in units of erg s$^{-1}$ cm$^{-2}$ {\AA}$^{-1}$,
and abscissa is the observed wavelength in angstrom.
The rest wavelength is given along the top axis.
The hatched regions indicate the place of telluric absorption.
}
\label{low_reso}
\end{figure}

\begin{figure}
  \begin{center}
    \FigureFile(90mm,110mm){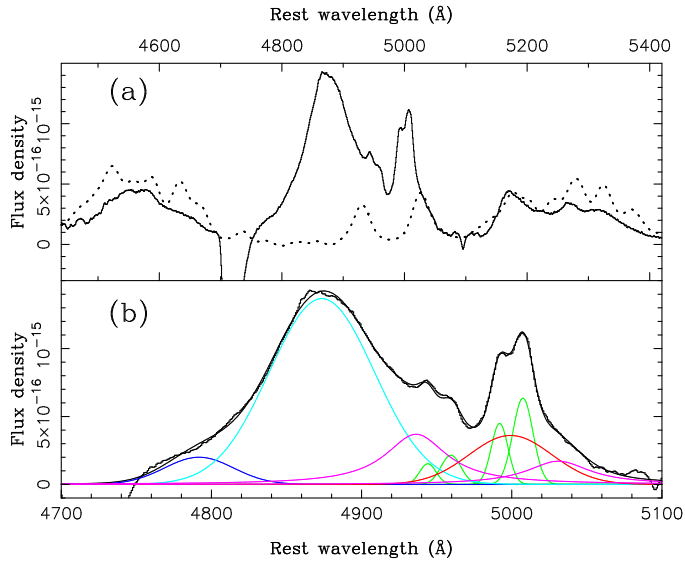}
  \end{center}
  \caption{The H$\beta$-[O~\emissiontype{III}] region of AKARI~J1757+5907.
(a) Continuum subtracted spectrum. 
The dotted line is the Fe~\emissiontype{II} template produced from I~Zw~1.
The Fe~\emissiontype{II} template is broadened and scaled at 5170 {\AA}.
Note that the Fe~\emissiontype{II} template is clearly different from Fe~\emissiontype{II} emission line of AKARI~J1757+5907 at 4600-4700 {\AA} and 5260-5400 {\AA}.
(b) Fit of the spectrum.
H$\beta$, [O~\emissiontype{III}] doublet, and Fe~\emissiontype{II} $\lambda\lambda4925, 5020$ are fitted with a three Gaussians (cyan, blue and red lines),
two sets of two Gaussians (green lines),
and two Lorentzians (magenta lines), respectively.}
\label{FeIIfit}
\end{figure}

Figure~\ref{low_reso} displays the low-resolution optical spectrum of 
AKARI~J1757+5907.
The spectrum shows strong absorption lines of Mg~\emissiontype{II}
and He~\emissiontype{I}* as well as emission lines of Mg~\emissiontype{II},
Fe~\emissiontype{II}, H$\gamma$, H$\beta$, [O~\emissiontype{III}].
In order to determine the systemic redshift of the quasar, we deblend the [O \emissiontype{III}] emission lines from the H$\beta$ and Fe~\emissiontype{II} emission lines.
As seen in figure~\ref{low_reso}, AKARI~J1757+5907 has strong Fe~\emissiontype{II} emission lines.
However, the Fe~\emissiontype{II} emission template around H$\beta$ derived from the spectrum of I~Zw~1 \citep{Aoki2005} 
is not a good match to the Fe~\emissiontype{II} emission in AKARI~J1757+5907
(figure~\ref{FeIIfit}a).
The intensity ratios among Fe~\emissiontype{II} emission lines are different between these objects.
Thus, we cannot use the Fe~\emissiontype{II} emission template derived form the spectrum of I~Zw~1.
\par
Instead we fit the spectrum between 7662 {\AA} and 8204 {\AA} with a 
combination of Fe~\emissiontype{II} $\lambda\lambda 4925, 5020$, H$\beta$, 
and [O~\emissiontype{III}] after subtraction of a power-law continuum.
This power-law continuum is constructed by fitting at 6526-6544 {\AA} and 8842-8891 {\AA}.
The H$\beta$ emission line is fitted with a combination of three Gaussians.
We fit the [O~\emissiontype{III}] doublet $\lambda\lambda 4960.3, 5008.2$ with two sets of two Gaussians.
The width and redshift are assumed to be the same for each set, 
and the intensity ratio is fixed to be 3.0.
Fe~\emissiontype{II} $\lambda\lambda 4925, 5020$ are modeled by two Lorentzians with the same width and redshift.
Their redshift is fixed to be the same as the strongest Gaussian component of the H$\beta$.
The result of fitting is shown in figure~\ref{FeIIfit}b.
We need a separate ``red wing'' of H$\beta$ to get a satisfactory fit.
This component may be [O~\emissiontype{III}] emission line.
The derived redshifts and FWHMs are tabulated in table \ref{tbl2}.
The FWHM is corrected for the instrumental broadening by using the simple
assumption:
$\rm{FWHM}_{true}=(\rm{FWHM}_{obs}^{2} - \rm{FWHM}_{inst}^{2})^{1/2}$,
where $\rm{FWHM}_{obs}$ is the observed FWHM of the line and
$\rm{FWHM}_{inst}$ is an instrumental FWHM (620 km s$^{-1}$).
The redshift of the red component of [O~\emissiontype{III}] is $0.6150\pm0.0001$.
The rest equivalent width of [O~\emissiontype{III}] including both components is
$9.9\pm0.3$ {\AA}.
This value is consistent with the [O~\emissiontype{III}] strength in majority ($>50$ \%) of 
Mg~\emissiontype{II} BAL quasars reported by \citet{2010ApJ...714..367Z}.
We also detect weak [O~\emissiontype{II}] emission at 6023.4 {\AA}, which corresponds to a redshift of 0.6155.
We thus adopt $0.61525\pm0.00025$ as the systemic redshift of the quasar.
The blue component of [O~\emissiontype{III}] is blueshifted by 980 km s$^{-1}$
relative to the systemic redshift.
\par

\section{Absorption lines}\label{coluumn_density}

\begin{figure}
  \begin{center}
    \FigureFile(90mm,110mm){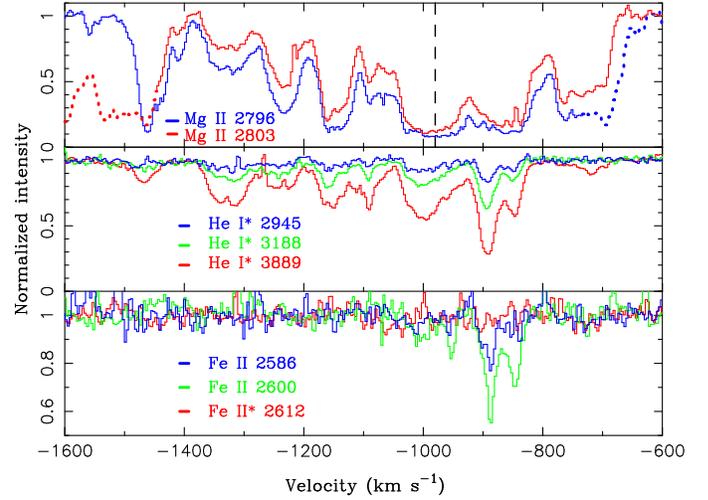}
  \end{center}
  \caption{Outflow troughs in AKARI~J1757+5907.
Ordinate is a normalized flux density,
and abscissa is velocity from the systemic redshift ($z=0.61525$).
Blended parts are denoted as dotted spectra.
The dashed vertical line indicates the position of the blue component of [O~\emissiontype{III}] emission line.
The velocity of that [O~\emissiontype{III}] corresponds to the trough at $\sim -1000$ km s$^{-1}$.}
\label{vel_plot}
\end{figure}

In the HDS spectrum of AKARI~J1757+5907, the absorption lines are not
heavily blended. The identification is thus a straightforward task.
We identify Mg~\emissiontype{II} $\lambda\lambda2976, 2803$,
He~\emissiontype{I}* $\lambda$2945, $\lambda$3188, $\lambda$3889,
Fe~\emissiontype{II}\ $\lambda$2600, $\lambda$2586 as well as weaker
absorption troughs from Mg~\emissiontype{I} $\lambda2852$,
He~\emissiontype{I}* $\lambda2829$, and Ca~\emissiontype{II}
$\lambda\lambda3934, 3969$. The strong absorption lines such as the
Mg~\emissiontype{II} doublet and He~\emissiontype{I}*$\lambda3889$
clearly show 9 distinct troughs, which span a velocity range from
$-660$ to $-1520$ km s$^{-1}$.  The trough at $-$1000 km s$^{-1}$ has
the same outflow velocity as the blue component of the
[O~\emissiontype{III}] emission line. We show the absorption troughs
from Mg~\emissiontype{II} $\lambda\lambda2976, 2803$,
He~\emissiontype{I}* $\lambda$2945, $\lambda$3188, $\lambda$3889, and
Fe~\emissiontype{II}\ $\lambda$2600, $\lambda$2586 in
figure~\ref{vel_plot}. We do not detect an Fe~\emissiontype{II}\
$\lambda2612$ absorption trough from the E=385 cm$^{-1}$ excited
level, which has the largest oscillator strength of the
Fe~\emissiontype{II}* lines from this energy level, in our spectral
range. We show the spectral region of Fe~\emissiontype{II}\
$\lambda2612$ in the lower panel of figure~\ref{vel_plot}.

\subsection{Column Density Determinations}

Both the Mg~\emissiontype{II} and He~\emissiontype{I}* troughs span 
the full velocity range from $-660$ to $-1520$ km s$^{-1}$ and appear in 
all 9 distinct troughs (see Figure~\ref{vel_plot}). There is self 
blending in the Mg~\emissiontype{II}\ troughs at the velocity 
extremes, which does not affect the majority of the troughs. The
He~\emissiontype{I}* and Fe~\emissiontype{II}\ troughs are free of 
any self blending. Of the 9 troughs, only three have corresponding 
absorption in Fe~\emissiontype{II}.
These are in the range of $-1050$ to $-800$ 
km s$^{-1}$ (see Figure~\ref{vel_plot}). Thus, we concentrate on this
velocity range for our measurements as the best photoionization 
constraints are achieved by contrasting He~\emissiontype{I}* and 
Fe~\emissiontype{II} column densities 
(see section~\ref{model}; \cite{Korista2008, Arav2010}).
\par
To determine the ionic column densities, we rebin the data to
10~km~s$^{-1}$ and use the velocity dependent apparent optical depth
(AOD), covering factor ($C(v)$), and power-law fitting methods from
Dunn et~al.\ (2010; methods 1, 2, and 3 in their subsection 3.2) across
the range of $-1050$ to $-800$ km s$^{-1}$. 
\par
The AOD methods assumes that the emission source is completely
and homogeneously covered by the absorber, so that the optical depth
($\tau(v)$) at a given velocity is related to the normalized intensity
via: $I(v)=\exp (-\tau(v))$.
The covering factor method assumes that
at a given velocity, a fraction $C(v)$ of the emission source is
covered with a constant value of optical depth, while the
rest of the source is uncovered.  In order to for both $\tau(v)$ and  $C(v)$ we use
use at least two absorption lines from the same energy level of the same ion, and solve for 
$I(v)_j=1-C(v)+C(v) \exp(-\tau(v)_j),$ where $I_j$ is the normalized intensity of the absorption due to the 
$j$ transition in the same energy level.  The ratio of different $\tau(v)_j$ are known from atomic physics and therefor the set of equations is solvable.
The power-law model assumes that the absorption gas inhomogeneously
covers the background source.  The optical depth is described by $
\tau_{v}(x)=\tau_{max}(v) x^{a},$ where $x$ is the spatial dimension in
the plane of the sky, $a$ is the power law distribution index and
$\tau_{max}$ is the highest value of $\tau$ at a given velocity.  In
case of the power law model, $\tau$ is averaged over the spatial
dimension x.
\par
In order to convert $\tau$ in each model to column density we use
      $$N(v) = 3.8 \times 10^{14} \frac{1}{f \lambda} \tau(v)  ({\rm cm}^{-2} {\rm km}^{-1} {\rm s}),$$
where $\lambda$ is the wavelength of the line in {\AA}, $f$ is the oscillator strength.
We use the oscillator
strengths from \citet{FW2006} for the Fe~\emissiontype{II} lines and 
values from the NIST Atomic Spectra Database (2010) for 
Mg~\emissiontype{II} and He~\emissiontype{I}*.
\par

\begin{figure}
  \begin{center}
  \FigureFile(90mm,110mm){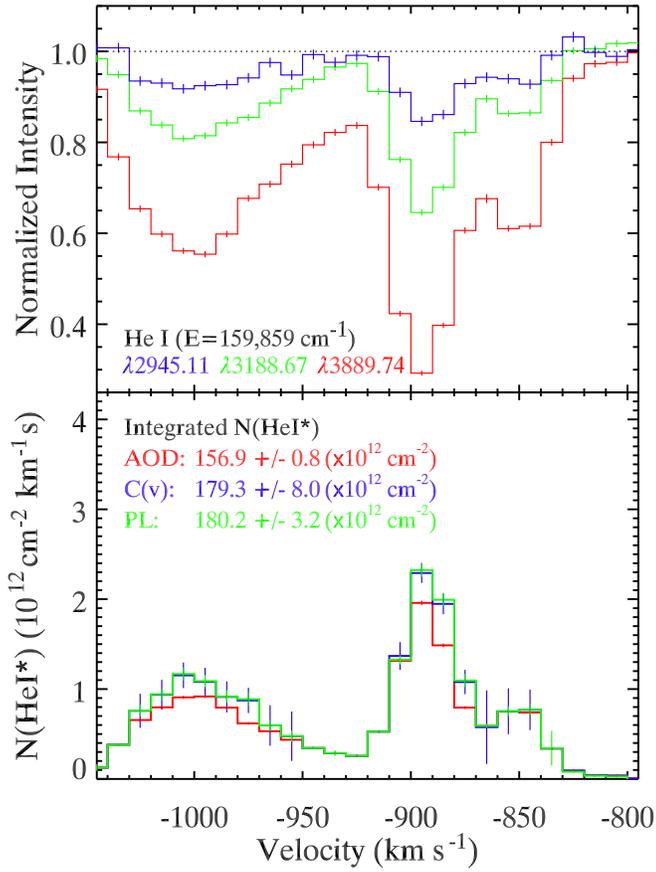}
\end{center}
  \caption {Top: Absorption troughs from He~\emissiontype{I} $\lambda$3890, 3190, 
and 2946 as red, green, and blue histograms, respectively. Vertical bars 
represent the statistical uncertainties in the residual intensity. 
Bottom: 
Velocity-dependent column density determinations for He~\emissiontype{I}* from the AOD 
(red histograms), $C(v)$ (blue histograms), and power-law (green 
histograms) solutions. The velocity-integrated He~\emissiontype{I}* column density 
values for the three methods and their associated statistical 
uncertainties are also listed.
}
 \label{hei}
\end{figure}

In Figure~\ref{hei}, we show the He~\emissiontype{I}* troughs and column density
determination from the three lines of He~\emissiontype{I}* present in the HDS 
spectrum. The He~\emissiontype{I}* lines are well separated, have no self blending, and 
we obtain consistent results with all three column density 
extraction methods. There are two velocity points where the $C(v)$ 
becomes nonphysical (i.e., negative or larger than
1.0), near $-930$ km s$^{-1}$ and at velocities lower than $-850$ 
km s$^{-1}$. This occurs because the two weaker lines are very 
shallow at these velocities and are thus consistent with the 
continuum level and dominated by the noise.
\par

\begin{figure}
\begin{center}
  \FigureFile(90mm,110mm){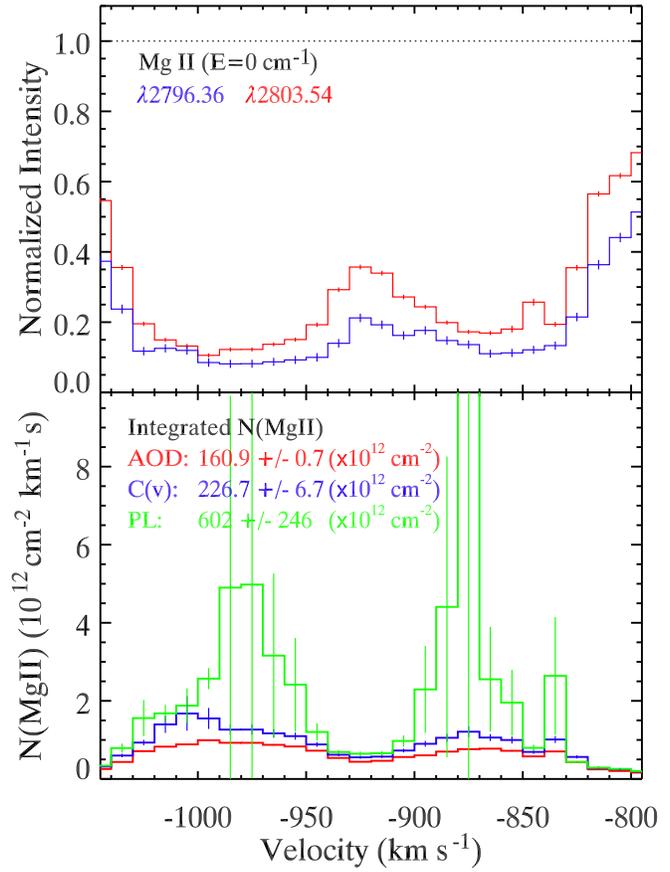}
  \end{center}
  \caption{Similar to Figure \ref{hei}. Top: This shows the trough
profiles for the Mg~\emissiontype{II} doublet lines $\lambda\lambda$2796, 2804 (blue 
and red histograms, respectively) and their associated statistical
uncertainties. 
Velocity-dependent column density determinations from the AOD 
(red histograms), $C(v)$ (blue histograms), and power-law (green 
histograms) solutions. The velocity-integrated column density 
values for the three methods and their associated statistical 
uncertainties are also listed.
}
  \label{mgii}
\end{figure}

We show in Figure \ref{mgii} the trough profiles and column density
results for Mg~\emissiontype{II}. Both the red and the blue doublet troughs are quite
deep and therefore their level of saturation is model dependent.  The
$C(v)$ solution suggests a small level of saturation (only 40\% larger
column density than the AOD estimate), while the power-law result is
three times higher.  This is an inherent feature of the absorption
models where in order to fit deep doublet troughs the power-law model
requires a much greater column density than the $C(v)$ solution (see 
the case of the O~\emissiontype{VI}\ doublet in the spectrum of Mrk~279, 
\cite{2005ApJ...620..665A}). Since we do not have data for troughs 
from additional Mg~\emissiontype{II} lines, we cannot determine which model 
is more physical. 
\par
There is a possibility that a narrow Mg~\emissiontype{II} emission line
fills the trough.
As already noted early in this section, the blue component of [O~\emissiontype{III}] 
has the velocity of -1000 km s$^{-1}$.
The Mg~\emissiontype{II} emission line from the same gas probably exists.
We estimate the flux of Mg~\emissiontype{II} emission line using the flux ratio of 
[O~\emissiontype{III}] $\lambda5008$ to Mg~\emissiontype{II} in Seyfert 2 galaxies, NGC~1068 \citep{Kra98} and Mrk~3 \citep{Collins2005}.
The ratios vary along the physical position in the galaxies between 0.03 and 0.15, and average of extinction corrected ratios are 0.09, and 0.07 in NGC~1068 and Mrk~3, respectively.
 We also assume a 1:2 ratio for the  Mg~\emissiontype{II} doublet, and
gaussians of the same width ($\sigma=205$ km s$^{-1}$) as the [O~\emissiontype{III}] emission line.
This width corresponds to 3.0 {\AA} at the Mg~\emissiontype{II} observed wavelength of 4500 {\AA}.
The HDS spectrum before normalization is scaled to the low-resolution spectrum.
The expected height of the Mg~\emissiontype{II} emission line will be $\sim 13$ \% and $\sim 4$\%
of the residual intensity at Mg~\emissiontype{II} $\lambda2796$ and
$\lambda 2803$ trough, respectively.
Thus, the real residual intensity may be smaller than the observed one.
Therefore, we conclude that the Mg~\emissiontype{II} 
column density can be much larger than the AOD or $C(v)$ determined values.
\par

\begin{figure}
  \begin{center}
  \FigureFile(90mm,110mm){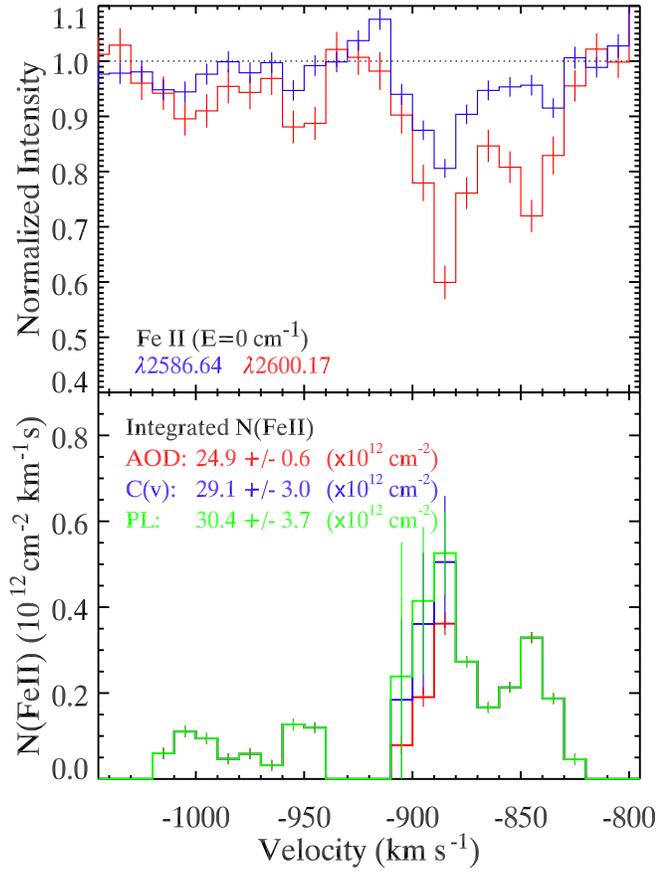}
  \end{center}
  \caption{Similar to Figures~\ref{hei} and \ref{mgii}. Top: The
trough profiles for the Fe~\emissiontype{II}\ $\lambda$2600 and 2587 lines as red and
blue histograms, respectively. Their associated statistical uncertainties
are shown as vertical slashes. Bottom: The AOD, $C(v)$, and power-law 
column densities plotted as a function of velocity and the respective 
integrated values in red, blue, and green, respectively.}
  \label{feii}
\end{figure}

Finally, we show the result for Fe~\emissiontype{II} troughs in Figure 
\ref{feii}. Unlike Mg~\emissiontype{II} and He~\emissiontype{I}*, the 
Fe~\emissiontype{II} troughs are both shallow and in a much lower
signal-to-noise region of the spectrum (towards the short wavelength end 
of the detector). Due to this, we find that the weaker Fe~\emissiontype{II} 
$\lambda$2587 line is only detected across three velocity bins. Using both 
the Fe~\emissiontype{II}\ $\lambda$2600 and the Fe~\emissiontype{II} 
$\lambda$2587 lines we calculate $N_{FeII}$ for both $C(v)$ and power-law 
methods for the three bins and include them in the integrated total. Due to 
a lack of detection of the $\lambda$2587 
line, we use the $\lambda$2600 line to calculate the column density from 
the AOD method for the remaining points. 
We also checked the 
column density for Fe~\emissiontype{II} using the covering factor of Mg~\emissiontype{II}. The column density 
calculated in this fashion changed only by $\sim$10\% as Mg~\emissiontype{II} nearly 
fully covers the source ($C(v)\approx$0.9) in this velocity range.

\subsection{Column Density Limits on Fe~\emissiontype{II}* Excited State Lines}

In order to help constrain the photoionization models in 
section~\ref{model}, we estimate the column density 
limits for the Fe~\emissiontype{II}* $\lambda$2612 and 
$\lambda$2757 lines from the 385 and 7955 cm$^{-1}$ energy 
levels, respectively. Neither line shows a detectable trough 
in the data. Therefore, we can use the trough profile of 
Fe~\emissiontype{II} $\lambda$2600 to determine the relative 
optical depth and column density of these two energy levels 
(see Section 3.3 of \cite{2010ApJ...709..611D}). We find upper 
limits on the ionic column densities of (3.7 and 0.5) 
$\times10^{12}$ cm$^{-2}$ for the 385
and 7955 cm$^{-1}$ energy levels, respectively.

\section{Determination of the Spectral Energy Distribution}\label{SED}

\begin{figure}
  \begin{center}
    \FigureFile(90mm,110mm){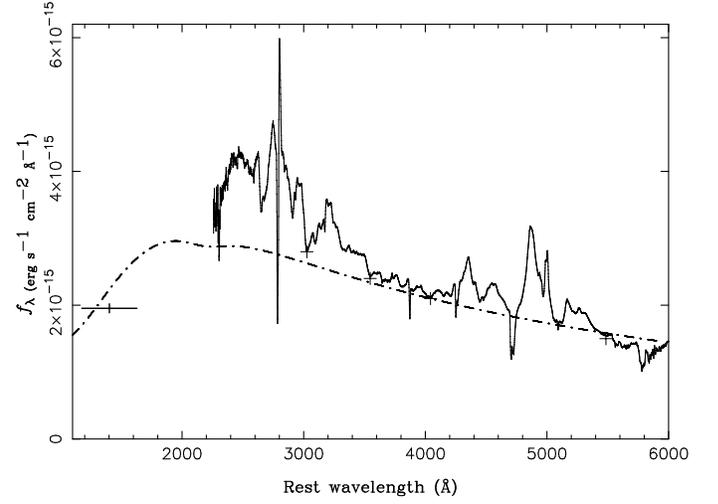}
  \end{center}
  \caption{Reddening of AKARI~J1757+5907.
  Ordinate is a flux density corrected for the Milky Way extinction in units of erg s$^{-1}$ cm$^{-2}$ {\AA}$^{-1}$
  and abscissa is a rest wavelength.
  The rest optical spectrum is shown with the GALEX photometry (the most left cross) and continuum measurements (four crosses) from our spectrum. 
  The power-law continuum reddened by the SMC-type extinction law with $E(B-V)=0.18$ is shown with a dot-dashed line.
 }
\label{deredMW2}
\end{figure}

Our FOCAS spectrophotometry data and the GALEX photometry clearly
show the flux drops at shorter wavelengths ($< 2000$ {\AA} in the 
rest frame) and indicate reddening by dust (figure~\ref{deredMW2}).
We must consider extinction for deriving the intrinsic spectral 
energy distribution (SED). First, we measured the continuum flux 
at four points where there are less Fe~\emissiontype{II} emission contaminates.
These four points and the GALEX photometry points are then shifted to the rest frame.
We fit a reddened power-law to the continuum points. 
We adopted the index of the power-law continuum ($\alpha$; $f_{\nu} \propto \nu^{\alpha}$) 
of the LBQS composite, -0.36 measured by \citet{VB01}.
We adopt the SMC-type extinction law.
The best fit gives us the color excess $E(B-V)$ of 0.18.
The best fit of the reddened power-law continuum is shown in figure~\ref{deredMW2}.
\par

To estimate the distance to the outflow from the central source ($R$),
we need to determine the flux of hydrogen ionizing photons that
irradiates the absorber (see equation [1] below). Using the
\citet{MF87} SED, reddened to match the observed spectrum, we find
that number of hydrogen ionizing photons emitted per second by the
reddened central source ($Q_H$) is 2.2$\times$10$^{57}$ photons
s$^{-1}$. Here we assumed that the reddening occur between the central
source and the outflow, as is the case where the photons are
attenuated by the edge of the putative AGN obscuring torus (see full
discussion in \cite{2010ApJ...709..611D}). This assumption will also
give us smaller values for the inferred $R$ and therefore conservative
lower limits for $\dot{M}_{out}$ and $\dot{E}_k$.  The $L_{Bol}$ for
the dereddened, intrinsic spectrum is $3.7 \times 10^{47}$ erg
s$^{-1}$.

\section{Photoionization Modeling}\label{model}
Through photoionization modeling, reliable measurements of
He~\emissiontype{I}* and Fe~\emissiontype{II}\ column densities
provide accurate constraints on the total hydrogen column density,
$N_H$, and the hydrogen ionization parameter,
\begin{equation}
U_H \equiv {Q_H\over {4\pi R^2 n_{H} c}},
\end{equation}
where $R$ is the distance from the central source, $n_{H}$ is the
total hydrogen number density, and $c$ is the speed of light. We use
version c08.00 of the spectral synthesis code Cloudy, last described
by \citet{Ferland1998}, to model a plane-parallel slab of gas with
constant hydrogen number density irradiated by a source continuum. We
focus on the kinematic components spanning a velocity range from
$-800$ km s$^{-1}$ to $-1050$ km s$^{-1}$ where we detect
Fe~\emissiontype{II}(0). In this velocity range,
measurements of upper limits on the 
Fe~\emissiontype{II}(E=385 cm$^{-1}$) and the 
Fe~\emissiontype{II}(E=7955 cm$^{-1}$) yield upper limits of electron number
density $n_{e}$, $10^{3.6}$ and $10^{3.8}$ cm$^{-3}$, respectively.
We adopt the conservative value of an upper limit of $n_{e} \leq
10^{3.8}$ cm$^{-3}$, and assume $n_{H} \approx n_{e}$, which is valid
within the ionized zone we are discussing.  While the electron number
density is well constrained from above, there are no diagnostics for a
lower limit on $n_{e}$ in the data.

We begin investigation of the parameter space by using Cloudy's
optimization mode to determine $N_H$ and $U_H$ for $n_{H}=10^{3.8}$
cm$^{-3}$. The parameters $N_H$ and $U_H$ are varied and ionic column
densities are computed for each set of parameters. Best fit values are
determined by $\chi^2$ minimization for given tolerances in the
measured ionic column densities.  We adopt the measured ionic column
densities determined by the partial covering method, and optimize to
 $N_{Fe~\emissiontype{II}}$ and
$N_{He~\emissiontype{I}*}$, since these are the more robust
measurements. The measured and model predicted column densities are
presented in table~\ref{coldens}.  For solar abundances and the MF87
SED, as implemented by Cloudy\footnote{This SED differs from the MF87
SED by the addition of a sub-millimeter break at 10 $\mu$m.}, we find
log $U_H=-2.15$ and log $N_H=20.82$ yield good fits to the column
densities of He~\emissiontype{I}* and Fe~\emissiontype{II}\, while
$N_{Mg~\emissiontype{II}}$ is overpredicted. However, as discussed in
section 4, the Mg~\emissiontype{II} troughs may be more saturated than
the partial covering model suggests. A hydrogen ionization front,
which we define as the position at which half of the total hydrogen is
neutral (approximated by $N_H=10^{23}U_H$), does not form in this
solution although we are very close to it with $\log (N_H/U_H)=22.96$.

Since the data do not provide a lower limit to the electron number
density, we find other valid solutions by reducing $n_{H}$. When the
hydrogen number density is reduced, the He~\emissiontype{I}*
population drops. This is due to the fact that He~\emissiontype{I}* is
populated by recombination of He~\emissiontype{II} and the number of
recombinations per unit time depends linearly on $n_e$. Therefore, $N_H$ 
must increase in order to provide enough He~\emissiontype{I}*
to be consistent with the measured value. Fe~\emissiontype{II}\ becomes dominant near the hydrogen
ionization front, while He~\emissiontype{I}* drops off drastically at
the front. Thus, solutions in this region of the slab have $U_H$ fixed
by $N_{He~\emissiontype{I}*}$ and the ratio $N_H/U_H$ fixed by
$N_{Fe~\emissiontype{II}}$. Due to the tight correlation of $N_H$ and
$U_H$, all valid solutions have a similar ratio laying (nearly) on a
straight line in the $N_H$-$U_H$ plane for number densities greater
than $\sim100$ cm$^{-3}$. At these densities, the slabs do not form a
hydrogen ionization front. As we go to densities lower than $\sim100$
cm$^{-3}$, a front forms, behind which Fe~\emissiontype{II}\ and
Mg~\emissiontype{II}\ increase linearly with $N_H$.

In order to determine what effects choice of SED may have on $N_H$ and
$U_H$, we compare results of the MF87 SED with results of a softer
SED. The soft SED has an optical to X-ray spectral index
$\alpha_{ox}=-1.5$ compared to the MF87 SED with $\alpha_{ox}=-1.4$
(with the convention $F_{\nu} = \nu^{\alpha}$) and was generated using
the Cloudy command agn 375000 -1.50 -0.125 -1.00, where the numbers
are the temperature of the UV bump, $\alpha_{ox}$, $\alpha_{uv}$,
$\alpha_{x}$, respectively. We find the resulting $N_H$ and $U_H$ are
nearly identical and conclude that changing the SED only affects the
energetics through $Q_H$, a finding consistent with the analysis of
QSO~1044+3656 reported in \citet{Arav2010}.

Another assumption in our models is solar abundances. To check the
sensitivity of our results to metallicity changes, we use the
abundances in table 2 of \citet{Ballero2008} for metallicity Z=4.23
with the MF87 SED and $\log n_{H}=3.8$. While helium abundances are
expected to increase with oxygen abundances (e.g., \citet{Olive1996}),
the amount of increase varies for different galaxies. Therefore, to be conservative,
we increase the helium abundance significantly to 15\% above solar. We find
that $\log U_H$ is approximately 0.02 dex lower and $\log N_H$ is
approximately 0.15 dex lower for the increased metallicity model with
$\log (N_H/U_H)=22.83$. We discuss the effect of these changes on the
energetics of the outflow in the next section.

\section{Energetics of the Outflow}\label{energetics}

Of particular interest for any outflow are its mass ($M_{out}$), the average mass flow rate ($\dot{M}_{out}$) and 
mechanical work output or kinetic luminosity ($\dot{E}_k$). Assuming the outflow is in a form of
a thin partial shell moving with a constant radial velocity ($v$), at a 
distance $R$ from the source, the mass of the outflow is:
\begin{equation}
M_{out} = 4 \pi \mu m_p \Omega R^2 N_H,
\end{equation}
where $N_H$ is the total column density of hydrogen, $m_p$ is the mass
of a proton, $\mu$=1.4 is the plasma's mean molecular weight per proton,
and $\Omega$ is the fraction of the shell occupied by the
outflow. 
The average mass flow rate is given by  dividing the outflowing mass by the dynamical time scale of the outflow $R/v$
(see full discussion in Arav et al 2010), therefore
\begin{equation} 
\dot{M}_{out}\sim \frac{M_{out}}{R/v} = 4 \pi \mu m_p \Omega R N_{H} v
\end{equation}
\begin{equation} 
{\rm and} \ \ \ \dot{E}_k = \frac{1}{2}\dot{M}_{out}v^2\sim 2 \pi \mu m_p \Omega R N_H v^3 .
\end{equation}

For the troughs we consider in AKARI~J1757+5907, the median velocity
of the system is $-970$ km s$^{-1}$. We assume that $\Omega$~=~0.2,
which is the percentage of quasars showing BALs in their spectrum (see
discussion in \cite{2010ApJ...709..611D}).

To determine the distance,
we solve equation (1) for $R$, which depends on $U_H$ and $n_{H}$.
The lack of Fe~\emissiontype{II}* detection yields an upper limit of
$n_H < 10^{3.8}$ cm$^{-3}$ (see Section~\ref{model}), and therefore,
as shown below, a lower limit on $R$.
 
\par For the range $10^{1.8}<n_{H} < 10^{3.8}$, our photoionization
solutions obey the relationships $N_H\propto U_H$ and $U_H\propto
n_{H}^{-\alpha}$, where $0.4<\alpha<1 $ ($\alpha$ decreases as $n_{H}$
increases). The first relationship arises from the requirement of
being close to a hydrogen ionization front, and the second is due to
the decreasing electron population at the He\emissiontype{I}* meta
stable level for lower number densities. Therefore, equation (1)
yields $R\propto n_{H}^{(\alpha-1)/2}$. We use the solar abundances
photoionization solution values from Section 6 ($\log U_{H}= -2.15$,
$\log N_H$ = 20.82 cm$^{-2}$ for $n_H=10^{3.8}$ cm$^{-3}$) and $Q_H$
from Section 5. Inserting these values into equation (1), provides a
lower limit on the distance, $R$ $>$ 3.7 kpc for $n_{H}=10^{3.8}$
cm$^{-3}$, which only increases to 6.6 kpc for
$n_{H}=10^{1.8}$(cm$^{-3}$) due to the weak dependence of $R$ on
$n_{H}$.  From equation (3) we observe that $\dot{M}_{out}$ and
$\dot{E}_k$ depend linearly on the product $RN_H$, which is
proportional to $n_{H}^{-(1+\alpha)/2}$. Therefore, the upper limit
for $n_{H}$ provides lower limit to $\dot{M}_{out}$ and $\dot{E}_k$.
Using the $N_H$ and $R$ derived for $n_{H}=10^{3.8}$ cm$^{-3}$, we
find lower limits of $\dot{M}_{out}$ $>$ 70 $\Omega_{0.2}$ $M_{\odot}$
yr$^{-1}$ and $\dot{E}_k$ $>$ 2.0 $\times$10$^{43}$ $\Omega_{0.2}$
ergs s$^{-1}$, where $\Omega_{0.2}\equiv\Omega/0.2$.  \par

In the previous section, we discussed changes in $N_H$ and $U_H$ due
to metallicity and SED changes. Using the \citet{Ballero2008} abundances
for Z/Z$_{\odot}$=4.23 reduces the mass flow rate by $\sim
30\%$. Changing to the soft SED described in the previous section
results in very small changes in $N_H$ and $U_H$, but $Q_H$ increases
by a factor of $\sim 2$, increasing the mass flow rate by a factor of
$\sim \sqrt{2}$.
\par

The mass of the black hole ($M_{BH}$) of AKARI~J1757+5907 is derived to
be $4.0 \times 10^{9} M_{\odot}$ based on the width of the H$\beta$ 
emission line of $\sigma=2160$ km s$^{-1}$ and the dereddened optical luminosity
$\lambda L(5100\mathrm{\AA})$ of $3.8 \times 10^{46}$ erg s$^{-1}$.
We use the formula in \citet{Bennert2010} based on the calibrations 
of broad-line region size-luminosity relation \citep{Bentz2006} and 
the virial coefficient taken from \citet{Onken2004}. The derived 
$\log L_{bol}/L_{Edd}=-0.13$ is used for calculation of mass accretion 
rate ($\dot{M}_{acc}$) based on the accretion disk model by 
\citet{Kawaguchi2003},
which takes into account the effects of electron scattering 
(opacity and disk Comptonization) and the relativistic effects.
The $\dot{M}_{acc}$ is 110 $M_{\odot}$ yr$^{-1}$,  which is similar to the lower limit of $\dot{M}_{out}$.
\par
We note that the $\Omega = 0.2$ we use is based the percentage of
quasars showing C~\emissiontype{IV} BALs.  
In a recent work, \citet{2010arXiv1004.0700D} have
shown that in the near infra-red surveys low ionization BALs (LoBALs) fraction is 4\%,
considerably higher than deduced from optical surveys (probably due to
obscuration effects in the optical band). A more appropriate
comparison in our case is to include somewhat narrower outflows with 1000
km s$^{-1} < \Delta v < 2000 $ km s$^{-1}$. For these LoBALs based on 
``Absorption index'' \citep{Trump06}, \citet{2010arXiv1004.0700D} 
find a 7.2\% fraction.  The frequency of He~\emissiontype{I}* 
outflows is
much less known.  The strongest He~\emissiontype{I}* line in the optical
($\lambda3889$) is shifted outside the optical range for objects where
we can detect C~\emissiontype{IV} $\lambda1550$ from the ground 
(e.g., z=1.5 for SDSS
spectra), so a meaningful census of these outflows is difficult to
come by.  Anecdotally, we find that in most cases where we detect
Fe~\emissiontype{II} absorption trough, we also detect He~\emissiontype{I}* troughs 
provided we
have a clear spectral coverage of the latter (e.g., \cite{2008ApJ...681..954A,Arav2010}).

We also point out that AKARI~J1757+5907 as well as QSO~2359-1241 do
not have measurable Fe~\emissiontype{II} absorption in their low
resolution spectra.  Also, their outflow velocities ($\sim 1200$ km
s$^{-1}$) and widths of troughs are moderate.  Quasars with similar
moderate outflows are more numerous than extreme FeLoBALs (e.g.,
SDSS~J0318-0600, \citet{Hal02}), and as we show here, can have similar
mass flow rate and kinetic luminosity as the more extreme ones.  This
fact suggests the mass flow rate and kinetic luminosity values found
here are more common among quasars, than judged by the rarity of
extreme FeLoBALs. If we assume as a conservative limit that the
$\Omega$ of the outflow seen in AKARI~J1757+5907 is similar to the
7.2\% LoBALs fraction found by \citet{2010arXiv1004.0700D} then
$\Omega_{0.2}=0.36$, which will reduce the values for mass flow rate
and kinetic luminosity accordingly.

\section{Discussion}\label{discussion}
In Table~\ref{outflow_tab} we show our $\dot{M}_{out}$ and
$\dot{E}_k$ determinations in quasar BAL outflows to date (the older,
less accurate, ones are shown in table 10 of Dunn et al 2010).  While
the lower limit on $\dot{E}_k$ for the AKARI J1757+5907 outflow is
rather low for AGN feedback purposes, we note that the corresponding
$\dot{M}_{out}$ value is large enough to yield a significant
contribution for the metal enrichment of the intra-cluster medium
around the parent galaxy (see Hallman and Arav 2010 [ApJ submitted])
\par

We also note that this moderate velocity outflow is similar to those
discovered in post-starburst galaxies at $z\sim0.6$
\citep{Tremonti2007}.  Those galaxies show an outflow with
Mg~\emissiontype{II} absorption and a velocity of 1000 km s$^{-1}$ .
Their stellar masses are as high as $(0.7-4.8) \times 10^{11}
M_{\odot}$ (calculated from stellar population synthesis modeling fit
to their spectra).  For comparison, the $M_{BH}$ of AKARI~J1757+5907
is $4.0 \times 10^{9} M_{\odot}$, therefore based on
$M_{BH}$-$M_{bulge}$ relation \citep{HR2004}, the bulge of its host
galaxy is estimated at $M_{bulge}1.8 \times 10^{12} M_{\odot}$. 
In addition, our derived $N_{H}$ is similar to the one
\citet{Tremonti2007} derived for outflow in the massive post-starburst
galaxies ($N_{H}=2\times 10^{20}$ cm s$^{-1}$).  They also pointed out
that AGNs probably exist in those post-starburst galaxies because they
get better fit to the spectra with a featureless blue power-law
continuum and they detected high-excited emission-lines such as
[O~\emissiontype{III}] and [Ne~\emissiontype{V}].  Thus, the outflows
seen in the massive post-starburst galaxies may be same phenomena as
outflows in quasars.  These facts may be one of the indications that
the outflow is a common phenomenon among massive galaxies.
\citet{Tremonti2007} had to make several assumptions to derive the physical
quantities of the outflow because high resolution spectroscopy for
such faint targets is difficult and non-detection of the important
diagnostic absorption lines from Fe~\emissiontype{II} (both ground and
excited states) and He~\emissiontype{I}*.  In contrast, our high-resolution
spectroscopy of BAL outflows in quasars, can constrain the
total hydrogen column density and the distance of outflowing gas from
the nucleus, which yield less model dependent estimates for $\dot{M}_{out}$ and
$\dot{E}_k$.

\section{Scientific gains with additional observations} \label{prospects}
\subsection{Additional Subaru HDS data}
We expect data with a much higher signal-to-noise ratio using Subaru/HDS
under good weather conditions.  A half hour exposure should have a
signal-to-noise ratio of 35 per pixel at 3800 {\AA} where
Fe~\emissiontype{II}* $\lambda2396$ is shifted to.  Our current upper
limit to Fe~\emissiontype{II}* $\lambda2612$ is derived from the data
with signal-to-noise ratio of 30 per pixel.  Fe~\emissiontype{II}*
$\lambda2396$ is 2.3 times stronger absorption than
Fe~\emissiontype{II}* $\lambda2612$.  With two hours ($4 \times 1800$
sec) of exposure time, we expect five times more strict upper limit of
the column density for the excited state of Fe~\emissiontype{II} (385
cm$^{-1}$).  Such an upper limit would translate to a hydrogen number
density of $n_{H}$ $< 10^{2.9}$ cm$^{-3}$. For this value of $n_{H}$
and our best model at that density ($\log U_H$ = -1.66 and $\log N_H$
= 21.32), we find $R$ $>$ 5.9~kpc, $\dot{E}_k$ $>$ 1.2
$\times$10$^{44}$ $\Omega_{0.2}$ ergs s$^{-1}$ and $\dot{M}_{out}$ $>$
340 $\Omega_{0.2}$ $M_{\odot}$ yr$^{-1}$.  It is of course possible
that we'll be able to detect Fe~\emissiontype{II}* $\lambda2396$
absorption, which will give us a determination of $n_{e}$ and
therefore measurements (instead of lower limits) for $\dot{M}_{out}$
and $\dot{E}_k$.

\par In this paper our analysis focused on the troughs for which we
can measure Fe~\emissiontype{II} column density.  The other six
troughs have only upper limits for Fe~\emissiontype{II} column density
given the current signal-to-noise ratio.  In order to study the
relationship between the troughs, and obtain the integrated values of
$\dot{M}_{out}$ and $\dot{E}_k$ for the full outflow, it is important
to measure Fe~\emissiontype{II} column density in all the
troughs. Combined with our other measurements, this will constrain the
ionization parameter and total hydrogen column density in each individual trough.  We detect 
He~\emissiontype{I}* absorption from 3 different lines in almost all the troughs
(figure 3).  The current data at He~\emissiontype{I}* $2945, 3188$
have a signal-to-noise-ratio of 65 per pixel.  In order to get similar
or better signal-to-noise ratio at Fe~\emissiontype{II} resonance
lines, $2383, 2586,$ and $2600$, we will need two additional hours of HDS
integration.

 \subsection{Imaging of outflowing gas}
The blue component of [O\emissiontype{III}] emission line has the same outflow velocity to the trough at $\sim 1000$ km s$^{-1}$.
Furthermore, the derived density and ionization parameter values for the trough are typical for the narrow-line regions of AGNs (see, e.g., \cite{Groves2004}).
The scale of 3.7 kpc in AKARI~J1757+5907 corresponds to \timeform{0.55''}.
The extended nebular gas associated with the outflow can be observed by the 
[O~\emissiontype{III}] emission line by $HST$ or the [S~\emissiontype{III}] $\lambda9533$ emission line by integral filed spectroscopy coupled with adaptive optics observations from large ground-based telescopes. 
Currently, the solid angle subtended by the outflow is the parameter with the largest
uncertainty.
Imaging the outflowing gas will directly determine this parameter.

\subsection{HST Cosmic Origins Spectrograph (COS) observations}
The near UV side of COS offers two important gains for the scientific
investigation of the outflow in AKARI~J1757+5907.  First, using the
G230L grating we can cover the full range of 1333--2800 {\AA}
(observed frame) with resolution of $\sim3000$ and obtain data with
S/N $\sim30$ per resolution element using a total of six {\it HST}
orbits.  Such data will allow us to connect the low ionization
absorption studied in this paper with the higher ionization phase seen in
Si~\emissiontype{IV}, C~\emissiontype{IV}, and N~\emissiontype{V}.  In
addition, this spectral range covers 3 pairs of
Si~\emissiontype{II}/Si~\emissiontype{II}* lines that can determine
number density considerably lower than is possible with the lines from
the Fe~\emissiontype{II}~E=385 cm$^{-1}$ lines (essentially the
critical density of the Si~\emissiontype{II}* level is an order of magnitude
lower than that of the Fe~\emissiontype{II}~E=385). In the unlikely
event that we will detect Si~\emissiontype{II}, but not
Si~\emissiontype{II}* absorption, which means that the gas density is
substantially below the critical density for the Si~\emissiontype{II}*
level, we will have C~\emissiontype{II}/C~\emissiontype{II}*
transitions covered that can determine the number density down to
$n_e\sim10$ cm$^{-3}$.  These diagnostics practically guarantee that
we will be able to determine $n_e$ and therefore the distance of the
outflow, as long as $R<100$ kpc.

In addition we can target the strongest pair of Si~\emissiontype{II}/Si~\emissiontype{II}* lines
(1260 {\AA}, 1265 {\AA}) for a higher resolution ($R\sim20,000$) in order to
obtain a fully resolved trough where we can obtain $n_e$ as a function
of velocity for all 9 outflow components.  With 6 HST orbits using the
COS 225M grating we can obtain high enough S/N to determine  $n_e$ for
25-50 resolved velocity points across the full width of the outflow
for a large dynamical range in $n_e$ ($100 < n_e < 3000$ cm$^{-3}$).  This
will allow for a sensitive tomography of the outflow and precise
determination of the distance to each kinematic component.

\bigskip
We are grateful to the staff of Subaru Telescope especially T.-S.~Pyo, and A.~Tajitsu, for their assistance during our HDS observations.
We thank T.~Hattori for taking FOCAS spectra kindly and providing to us.
We also thank Jong-Hak Woo, T. Kawaguchi and the referee for their helpful comments.
This work was done when K.~A. and K.~T.~K. were staying at the Physics Department of Virginia Tech in summer 2010.
K.~A. and K.~T.~K. thank the staff there for hospitality and support during their stay.
We acknowledge support form NSF grant AST 0837880.
This publication makes use of data products from the Two Micron All Sky Survey, 
which is a joint project of the University of Massachusetts and the Infrared Processing and Analysis Center/California Institute of Technology, funded by the National Aeronautics and Space Administration and the National Science Foundation.

\clearpage

\clearpage 


\begin{table}
  \caption{Photometry of AKARI~J1757+5907.}\label{tbl1}
  \begin{center}
\begin{tabular}{ccc}
\hline
Source & Band & flux density/magnitude\\
\hline
GALEX & NUV &    $230.55\pm8.03 \mu$Jy$^{*}$ \\
POSS-I& $g$   & 15.1 mag.\\
POSS-I& $r$   & 15.0 mag.\\
POSS-II& $g$   & 15.2 mag.\\
POSS-II& $r$   & 15.3 mag.\\
POSS-II& $i$   & 15.1 mag.\\
2MASS & $J$ &     $13.728\pm0.021$  mag.    \\
2MASS & $H$ &     $13.074\pm0.024$ mag.     \\
2MASS & $K_{s}$ & $12.344\pm0.023$ mag.   \\
\hline
\multicolumn{3}{@{}l@{}}{\hbox to 0pt{\parbox{85mm}{\footnotesize
\par\noindent
\footnotemark[$*$]This value corresponds to $(1.345\pm0.047) \times 10^{-15}$ erg s$^{-1}$ cm$^{-2}$ {\AA}$^{-1}$ at 2267 {\AA}.
}\hss}}
\end{tabular}
\end{center}
\end{table}

\begin{table}
  \caption{Properties of emission lines of AKARI~J1757+5907.}\label{tbl2}
  \begin{center}
\begin{tabular}{cccc}
\hline
     & $\lambda_{\rm obs}$ & & ${\rm FWHM}_{\rm true}$ \\
Line & {\AA}             & $z$ & km s$^{-1}$ \\
\hline
H$\beta$        & 7872.0 & 0.6189 & 5080 \\
{[O~\emissiontype{III}]} 5008.24& 8088.3 & 0.6150 & 710 \\
{[O~\emissiontype{III}]} 5008.24& 8063.3 & 0.6100 & 480 \\
\hline
\end{tabular}
\end{center}
\end{table}

\begin{table}
\caption{Measured and model predicted column densities} \label{coldens}
\begin{center}
\begin{tabular}{lccccccc}
\hline
\multicolumn{5}{c}{Measurement Method / Model parameter} & $N_{He~\emissiontype{I}*}$$^*$ & $N_{Fe~{II}(0)}$$^*$ & $N_{Mg~{II}}$$^*$\\
\hline
\multicolumn{8}{c}{Measurements}\\
\multicolumn{5}{l}{Apparent Optical Depth} & 156.9 $\pm$0.8 & 24.9 $\pm$0.6 & 160.9 $\pm$0.7\\
\multicolumn{5}{l}{Covering factor} & 179.3 $\pm$8.0 & 29.1 $\pm$3.0 & 226.7 $\pm$6.7\\
\multicolumn{5}{l}{Power Law} & 180.2 $\pm$3.2 & 30.4 $\pm$3.7 & 602 $\pm$246\\
\hline
\multicolumn{8}{c}{Models}\\
SED & Z/Z$_{\odot}$ & $\log N_{H}$ & $\log U_{H}$ & $\log n_{H}$ & & &\\
MF87 & 1.00 & 20.82 & -2.15 & 3.8 &  176.5 & 30.9 & 1341.5\\
Soft & 1.00 & 20.81 & -2.15 & 3.8 & 180.5 & 32.4 & 1364.6\\
MF87 & 4.23 & 20.66 & -2.17 & 3.8 &  172.3 & 31.2 & 1815.5\\
\hline
\multicolumn{8}{@{}l@{}}{\hbox to 0pt{\parbox{85mm}{\footnotesize
\par\noindent
\footnotemark[$*$]in units of $10^{12}$ cm$^{-2}$
}\hss}}
\end{tabular}
\end{center}
\end{table}


\begin{table}
  \caption{Properties of Measured Outflows to Date}\label{outflow_tab}
  \begin{center}
\begin{tabular}{lccccccc}
\hline
Object & log $L_{Bol}$ & $R$ & log $N_H$ & log $U_H$ & log $\dot{E_k}^*$ & 
${\dot{M}_{out}}^*$ & Reference$^\dagger$ \\
 & (ergs s$^{-1}$) & (kpc) & (cm$^{-2}$) & & (ergs s$^{-1}$) &($M_{\odot}$ yr$^{-1}$) &\\
\hline
AKARI~J1757+5907    & 47.57 & $> 3.7$  & $> 20.82$ & $> -2.15$ & $> 43.30$ & $> 70$   & 1 \\
QSO 1044+3656     & 46.84 & 1.7 $\pm$ 0.4        & 20.84 $\pm$ 0.10 & $-$2.19 $\pm$ 0.10& 44.81 $^{+0.09}_{-0.11}$ & 120 $\pm$ 25        & 2 \\
QSO 2359$-$1241   & 47.67 & 3.2 $^{+1.8}_{-1.1}$ & 20.56 $\pm$ 0.15 & $-$2.40 $\pm$ 0.15& 43.36 $\pm$ 0.27         & 90 $^{+35}_{-20}$ & 3 \\
SDSS J0838+2955   & 47.53 & 3.3 $^{+1.5}_{-1.0}$ & 20.80 $\pm$ 0.28 & $-$1.95 $\pm$ 0.21& 45.35 $^{+0.23}_{-0.22}$ & 300 $^{+210}_{-120}$ & 4 \\
SDSS J0318$-$0600 & 47.69 & 5.9 $\pm$ 0.4        & 19.90 $\pm$ 0.17 & $-$3.08 $\pm$ 0.05& 44.55 $^{+0.10}_{-0.15}$ & 60 $\pm$ 20  & 5 \\
\hline
\multicolumn{8}{@{}l@{}}{\hbox to 0pt{\parbox{180mm}{\footnotesize
\par\noindent
\footnotemark[$*$]Calculated using equation (3) assuming $\Omega=0.2$.  The values for the last three objects are half of those founds in the reference due to the use of an improved estimate for $M_{\odot}$ and $\dot{E_k}$ given by  equation (3), over those given by equations (9) and (10) in 
Dunn et al. (2010).
\\
\footnotemark[$\dagger$] 1-This Work, 2-Arav et al. (2010), 3-Korista et al. (2008) \& Bautista et al. (2010), 
4-Moe et al. (2009), 5-Dunn et al. (2010)
}\hss}}
\end{tabular}
\end{center}
\end{table}


\begin{thebibliography}{}
\bibitem[Aoki, Kawaguchi, \& Ohta(2005)]{Aoki2005} Aoki,~K., Kawaguchi,~T., \& Ohta,~K. 2005, \apj, 618, 601
\bibitem[Arav et al.(2001)]{Arav2001} Arav,~N., Brotherton,~M.~S., Becker,~R.~H., Gregg,~M.~D., White,~R.~L., Price,~T., \& Hack,~W. 2001, \apj, 546, 140
\bibitem[Arav et al.(2005)]{2005ApJ...620..665A} Arav,~N., Kaastra,~J., Kriss,~G.~A., Korista,~K.~T., Gabel,~J., \& Proga,~D. 2005 \apj, 620, 665
\bibitem[Arav et al.(2008)]{2008ApJ...681..954A} Arav,~N., Moe,~M., Costantini,~E., Korista,~K.~T., Benn, C., \& Ellison,~S. 2008, \apj, 681, 954
\bibitem[Arav et al.(2010)]{Arav2010} Arav,~N., Dunn,~J.~P., Korista,~K.~T., Edmonds,~D., Gonz\'{a}lez-Serrano,~J.~I., Benn,~C., \& Jim\'{e}nez-Luj\'{a}n,~F. 2010, submitting to \apj
\bibitem[Ballero et al.(2008)]{Ballero2008} Ballero,~S.~K., Matteucci,~F., Ciotti,~L., Calura,~F., \& Padovani,~P. 2008, \aap, 478, 335 
\bibitem[Barlow \& Sargent(1997)]{1997AJ....113..136B} Barlow,~T.~A. \& Sargent,~W.~L.~W. 1997, \aj, 113, 136
\bibitem[Bautista et al.(2010)]{Bautista2010} Bautista,~M.~A., Dunn,~J.~P., Arav,~N., Korista,~K.~T., Moe,~M., \& Benn,~C. 2010, \apj, 713, 25
\bibitem[Brotherton et al.(2001)]{Brotherton2001} Brotherton,~M. ~S., Arav,~N., Becker,~R.~H., Tran,~H.~D., Gregg,~M.~D., White,~R.~L., Laurent-Muehleisen,~S.~A., \& Hack, W. et al. 2001, \apj, 546, 134
\bibitem[Bennert et al.(2010)]{Bennert2010}Bennert,~V.~N., Treu,~T., Woo,~J.-H., Malkan,~M.~A., Auger,~M.~W., Gallagher,~S., \& Blandford,~R.~D. 2010, \apj, 708, 1507
\bibitem[Bentz et al.(2006)]{Bentz2006}Bentz,~M.~C., Peterson,~B.~M., Pogge,~R.~W., Vestergaard,~M., \& Onken,~C.~A. 2006, \apj, 644, 133
\bibitem[Br\"{u}ggen \& Scannapieco(2009)]{2009MNRAS.398..548B} Br\"{u}ggen,~M. \& Scannapieco,~E. 2009, \mnras, 398, 548
\bibitem[Ciotti et al.(2010)]{2010ApJ...717..708C} Ciotti,~L., Ostriker,~J.~P., \& Proga,~D. 2010, \apj, 717, 708
\bibitem[Collins et al.(2005)]{Collins2005} Collins,~N.~R., Kraemer,~S.~B., Crenshaw,~D.~M., Ruiz,~J., Deo,~R., \& Bruhweiler,~F.~C. 2005, \apj, 619, 116
\bibitem[de Kool et al.(2001)]{2001ApJ...548..609D} de Kool, M., Arav,~N., Becker,~R.~H., Gregg,~M.~D., White,~R.~L., Laurent-Muehleisen,~S.~A., Price,~T., \& Korista,~K.~T. 2001, \apj, 548, 609 
\bibitem[Condon et al.(1998)]{1998AJ....115.1693C}Condon,~J.~J., Cotton,~W.~D., Greisen,~E.~W., Yin,~Q.~F., Perley,~R.~A., Taylor,~G.~B., \& Broderick,~J.~J. 1998, \aj, 115, 1693
\bibitem[Dai et al.(2010)]{2010arXiv1004.0700D} {Dai}, X., {Shankar}, F. and {Sivakoff}, G.~R., arXiv:1004.0700
\bibitem[de Kool et al.(2002)]{2002ApJ...570..514D} de Kool,~M., Becker,~R.~H., Arav,~N., Gregg,~M.~D., White,~R.~L. 2002, \apj, 570, 514
\bibitem[Di Matteo et al.(2005)]{2005Natur.433..604D} Di Matteo,~T., Springel,~V., \& Hernquist,~L. 2005, \nat,  433, 604
\bibitem[Dunn et al.(2010)]{2010ApJ...709..611D}Dunn, J.~P. et al. 2010, \apj, 709, 611
\bibitem[Ferland et al.(1998)]{Ferland1998}Ferland,~G.~J., Korista,~K.~T., Verner,~D.~A., Ferguson,~J.~W., Kingdon,~J.~B., Verner,~E.~M. 1998, \pasp, 110, 761
\bibitem[Fuhr \& Wiese(2006)]{FW2006}Fuhr,~J., \& Wiese,~W. 2006, J. Phys. Chem. Ref. Data, 35, 1669
\bibitem[Ganguly \& Brotherton(2008)]{2008ApJ...672..102G}Ganguly,~R., \& Brotherton,~M.~S. 2008, \apj, 672, 102
\bibitem[Groves, Dopita, Sutherland(2004)]{Groves2004} Groves,~B.~A., Dopita,~M.~A., Sutherland,~R.~S. 2004, \apjs, 153, 9
\bibitem[Hall et al.(2002)]{Hal02} Hall, P. B., et al. 2002, \apjs, 141, 267
\bibitem[Hamann et al.(2001)]{2001ApJ...550..142H} Hamann,~F.~W., Barlow,~T.~A., Chaffee,~F.~C., Foltz,~C.~B., \& Weymann,~R.~J. 2001, \apj, 550, 142
\bibitem[H\"{a}ring \& Rix(2004)]{HR2004}H\"{a}ring,~N.,\& Rix,~H.-W. 2004, \apjl, 604, L89
\bibitem[Hewett \& Foltz(2003)]{2003AJ....125.1784H}Hewett,~P.~C. \& Foltz,~C.~B. 2003, \aj, 125, 1784
\bibitem[Hopkins et al.(2008)]{2008ApJS..175..356H} Hopkins,~P.~F., Hernquist,~L., Cox,~T.~J., \& Ker\v{e}s,~D. 2008, \apjs, 175, 356
\bibitem[Ishihara et al.(2010)]{Ishihara2010} Ishihara,~D., et al. 2010, \aap, 514, A1
\bibitem[Iye et al.(2004)]{Iye04} Iye, M., et al. 2004, \pasj, 56, 381
\bibitem[Kashikawa et al.(2002)]{Kashik02} Kashikawa, N., et al. 2002, \pasj, 54, 819
\bibitem[Kawaguchi(2003)]{Kawaguchi2003}Kawaguchi,~T. 2003, \apj, 593, 69
\bibitem[Kellermann et al.(1989)]{1989AJ.....98.1195K}Kellermann,~K.~I., Sramek,~R., Schmidt,~M., Shaffer,~D.~B., \& Green,~R. 1989, \aj, 98, 1195
\bibitem[Knigge et al.(2008)]{2008MNRAS.386.1426K} Knigge,~C., Scaring,~S., Goad,~M.~R., \& Cottis,~C.~E. 2008, \mnras, 386, 1426
\bibitem[Korista et al.(2008)]{Korista2008} Korista,~K.~T., Bautista,~M.,~A., Arav,~N., Moe,~M., Costantini,~E., \& Benn,~C. 2008, \apj, 688, 108
\bibitem[Kraemer et al.(1998)]{Kra98} Kraemer,~S.~B., Ruiz,~J.~R., \& Crenshaw,~D.~M. 1998, \apj, 508, 232
\bibitem[Mathews \& Ferland(1987)]{MF87} Mathews,~W.~G. \& Ferland,~G.~J. 1987, \apj, 323, 456
\bibitem[Moe et al.(2009)]{2009ApJ...706..525M} Moe,~M., Arav,~N., Bautista,~M.,~A., \& Korista,~K.~T. 2009, \apj, 706, 525 
\bibitem[Moll et al.(2007)]{2007AA...463..513M}Moll,~R., et al., 2007, \aap, 463, 513
\bibitem[Monet et al.(2003)]{Monet03} Monet,~D.,~G., et al. 2003, \aj, 125, 984
\bibitem[Murakami et al.(2007)]{Murakami07} Murakami, H., et al., 2007, \pasj, 59, 369
\bibitem[Noguchi et al.(2002)]{Noguchi2002} Noguchi, K., et al. 2002, \pasj, 54, 855
\bibitem[Olive \& Scully(1996)]{Olive1996}Olive,~K.~A. \& Scully,~S.~T. 1996, International Journal of Modern Physics A, 11, 409
\bibitem[Onken et al.(2004)]{Onken2004}Onken~C.~A., Ferrarese,~L., Merritt,~D., Peterson,~B.~M., Pogge,~R.~W., Vestergaard,~M., \& Wandel~A. 2004, \apj, 615, 645
\bibitem[Schlegel, Finkbeiner, \& Davis(1998)]{SFD98} Schlegel,~D.~J.,Finkbeiner,~D.~P., \& Davis,~M. 1998, \apj, 500, 525 
\bibitem[Skrutskie et al.(2006)]{skrutskie06} Skrutskie, M.~F., et al.\ 2006, \aj, 131, 1163
\bibitem[Tremonti, Moustakas, Diamond-Stanic(2007)]{Tremonti2007} Tremonti,~C.~A., Moustakas,~J., \& Diamond-Stanic,~A.~M. 2007, \apjl, 663, L77
\bibitem[Trump et al.(2006)]{Trump06} Trump,~J. et al. 2006, \apjs, 165,1
\bibitem[Vanden Berk et al.(2001)]{VB01} Vanden Berk,~D.~E. et al., 2001, \aj, 122, 549
\bibitem[Wampler,~Chugai,~Petitjean(1995)]{1995ApJ...443..586W}Wampler,~E.~J., Chugai,~N.~N., \& Petitjean,~P. 1995, \apj, 443, 586
\bibitem[Weymann et al.(1991)]{1991ApJ...373...23W} Weymann,~R.~J., Morris,~S.~L., Foltz,~C.~B., \& Hewett,~P.~C. 1991, \apj, 373, 23
\bibitem[Zhang et al.(2010)]{2010ApJ...714..367Z} {Zhang}, S., {Wang}, {T.-G.}, {Wang}, H., {Zhou}, H., {Dong}, {X.-B.} \& {Wang}, {J.-G.}, \apj, 714, 367
\end{thebibliography}
\end{document}